\documentclass[final]{aipproc}
\layoutstyle{6x9}

\usepackage{graphicx}

\begin{document}
  \DeclareGraphicsExtensions{.eps}
  
  \title{Recent STAR Results from Charged Pion Production in Polarized pp
  Collisions at $\sqrt{s} = 200 GeV$ at RHIC}
  
  
  \author{Adam Kocoloski (for the STAR Collaboration)}{
    address={Massachusetts Institute of Technology\linebreak
    77 Massachusetts Ave., Cambridge, MA  02139}
  }
  
  \begin{abstract}
    The STAR experiment at RHIC measures the longitudinal double-spin asymmetry
    $A_{LL}$ for a variety of final states in collisions of longitudinally
    polarized protons to constrain the polarized gluon distribution in the
    proton. Asymmetries for mid-rapidity charged pion production benefit from
    large cross-sections and the excellent tracking and particle identification
    capabilities of the STAR Time Projection Chamber. This contribution presents
    a measurement of the ratio of cross sections for inclusive $\pi^{-}$ and
    $\pi^{+}$ production using data collected in 2005, as well as a
    new measurement of $A_{LL}$ for charged pions opposite a jet obtained
    from the 2006 RHIC run.
  \end{abstract}
  
  \maketitle
  
  Collisions of polarized protons at the Relativistic Heavy Ion Collider (RHIC)
  offer a window into the spin composition of the proton through measurements of
  longitudinal double-spin asymmetries in a variety of final states
  \cite{Bunce:2000uv}:
  
  \begin{equation}
    A_{LL} = \sum_{ABC} \frac{\Delta f_{A} \otimes \Delta f_{B} \otimes \hat
    a_{LL} \otimes \sigma_{AB \rightarrow CX} \otimes D_{C}^{\pi^{+}}} {f_{A}
    \otimes f_{B} \otimes \sigma_{AB \rightarrow CX} \otimes D_{C}^{\pi^{+}}}.
  \end{equation}
  
  Charged pion production at STAR \cite{Ackermann:2002ad} is dominated by gg and
  qg scattering, providing sensitivity to the polarized gluon distribution
  $\Delta g(x,Q^{2})$ over a restricted kinematic region of $0.03 < x < 0.3$.
  Furthermore, we know that a $\pi^{+}$ is more likely to have fragmented from a
  u quark ("favored" fragmentation) than a d quark ("disfavored" fragmentation).
  The converse is true for $\pi^{-}$, and the ratio of favored to disfavored
  fragmentation grows as the fraction of the parton momentum carried by the pion
  increases. This feature makes $A_{LL}(\pi^{+})$ a particularly attractive
  measurement, since its analyzing power is magnified by the significant u quark
  polarization in the kinematic range where qg scattering dominates the pion
  production cross section.
  
  One can write $A_{LL}$ in terms of experimental quantities as
  \begin{equation}A_{LL} = \frac{1}{P_{1}P_{2}} \frac{N^{++} - RN^{+-}}{N^{++} +
  RN^{++}},\end{equation} where $P_{1,2}$ are the polarizations of the colliding
  proton beams, $N^{++}$ and $N^{+-}$ are the identified particle yields when
  the proton helicities are aligned and anti-aligned, and R is the ratio of the
  beam luminosities in the two helicity configurations. The beam polarizations
  at RHIC are measured every few hours using a high-statistics Coulomb Nuclear
  Interference (CNI) polarimeter \cite{Jinnouchi:2004up}, and the analyzing
  power of the CNI polarimeter is normalized using a gas jet polarimeter
  \cite{Okada:2006dd}. The spin-dependent relative luminosities are measured at
  STAR using the Beam Beam Counters (BBCs), segmented scintillator annuli that
  provide full azimuthal coverage and span $3 < |\eta| < 5$ in rapidity .
  Coincident signals in the two BBCs define STAR's $pp$ minimum-bias trigger
  condition; this approach samples 87\% of the non-singly diffractive scattering
  cross section \cite{Adams:2003kv}. The BBCs can also be used to measure
  residual transverse beam polarization, which manifests as an azimuthal
  asymmetry in the scintillator tile counts.
  
  The minimum-bias trigger condition is necessary but not sufficient to select
  events for this analysis. Also required is an electromagnetic energy deposit
  in the Barrel Electro-Magnetic Calorimeter (BEMC) which enhances STAR's
  sampling of hard scattering events. This jet patch (JP) trigger condition
  splits the BEMC into fixed patches with an extent of $\Delta\eta \times
  \Delta\phi = 1.0 \times 1.0$ and looks for summed energy above a threshold in
  at least one patch. In the 2005 RHIC run the JP trigger was implemented for $0
  < \eta < 1$ with a primary threshold of 6.5 GeV. In 2006 the BEMC was fully
  commissioned and the JP trigger acceptance doubled to $|\eta| < 1.0$ with a
  threshold of 8.3 GeV. The BEMC has a depth of $\sim$1 hadronic interaction
  length, so charged pions often leave only a MIP signal of 264 MeV
  \cite{Cormier:2001vu} in the calorimeter.
  
  STAR reconstructs and identifies charged pions using a large Time Projection
  Chamber (TPC) situated inside a 0.5T magnetic field. The TPC can measure the
  transverse momenta of charged particles up to 20 GeV/c, and measurements of
  the energy loss along the track allow for particle identification across a
  wide range of momenta. The dE/dx and momentum of a track are nominally
  compared to a Bichsel parameterization of charged particle interactions in the
  TPC gas volume to obtain its PID; corrections to the Bichsel curves using
  particles identified via other means (electrons from the BEMC, pions and
  protons from strange decays) can improve the accuracy of dE/dx-based PID and
  allow for pion identification up to 15 GeV/c \cite{Xu:2008}.
  
\begin{figure}
    \includegraphics[width=0.69\textwidth]{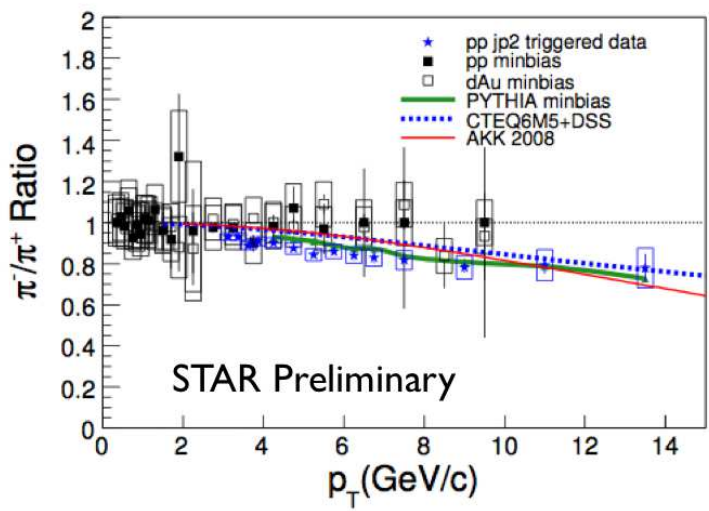}
    \caption{Ratio of invariant yields for charged pion production versus
    $p_{T}$. Statistical uncertainties are represented by vertical lines and
    systematics by the open boxes. The data are compared to predictions from
    \textsc{pythia} \cite{Sjostrand:2006za} and NLO analyses incorporating
    modern fragmentation functions \cite{deFlorian:2007aj, Kniehl:2008et}.}
    \label{fig:pionratio}
  \end{figure}

  Figure \ref{fig:pionratio} shows the ratio of cross sections for inclusive
  $\pi^{-}$ and $\pi^{+}$ production as a function of $p_{T}$, using data taken
  by STAR during the 2005 run. Both cross sections are separately consistent
  with NLO predictions \cite{Xu:hq2008}. The ratio clearly diverges from unity
  at high transverse momenta, and this divergence is modeled well by NLO pQCD
  predictions incorporating modern fragmentation functions from the DSS
  \cite{deFlorian:2007aj} and AKK \cite{Kniehl:2008et} collaborations.
  
  The data sample obtained by STAR during the 2006 RHIC run offers a factor of 6
  increase in the $A_{LL}$ figure of merit ($P^4 \ast \mathcal{L}$) and a
  doubling of the jet patch trigger acceptance relative to 2005. However, the
  increased thresholds used in the 2006 version of the JP trigger lead to a
  significant fragmentation bias for jets that fire the trigger. Rather than
  fold this bias into a systematic uncertainty on $A_{LL}$, we restricted the
  2006 charged pion $A_{LL}$ to pions opposite a trigger jet ($\Delta \phi >
  2.0$). Furthermore, we plot $A_{LL}$ not against pion $p_{T}$ but against the
  ratio of pion $p_{T}$ and trigger jet $p_{T}$. This ratio serves as a
  surrogate for the fragmentation variable $z$ and, along with a restriction on
  the trigger jet $p_{T}$, allows the analysis to identify a region of phase
  space where the effect of $\Delta g(x)$ on $A_{LL}$ is magnified by the large
  u quark polarization.
  
  Events are selected for the $A_{LL}$ analysis if they satisfy the BBC and JP
  trigger conditions outlined above. Jet reconstruction proceeds by clustering
  TPC tracks and BEMC towers using the midpoint-cone algorithm with a cone
  radius of 0.7. A jet is identified as a trigger jet if its axis points within
  36\textdegree~of the center of a JP above threshold. STAR also requires that
  the fraction of a jet's energy from the BEMC is less than 0.92 (to cut down on
  false jets formed out of beam background), and in this analysis we select jets
  with $9 < p_{T} < 25$ GeV/c to enhance our sample of qg scattering. Charged
  pions opposite the trigger jet are selected from a high-quality sample of TPC
  tracks with $p_{T} > 2$ GeV/c and $|\eta| < 1.0$, and are identified with a
  purity of better than 90\% using dE/dx.
  
\begin{figure}
    \includegraphics[width=\textwidth]
      {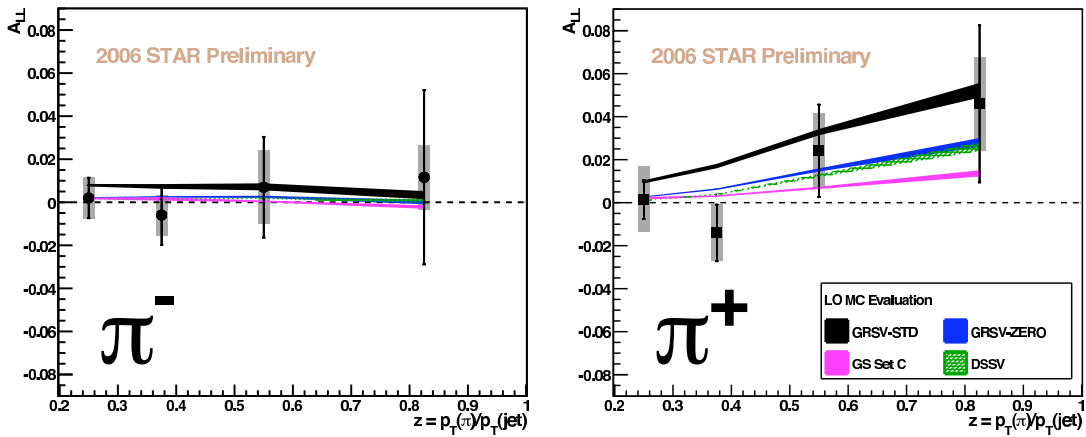}
    \caption{$A_{LL}$ for charged pions opposite a trigger jet. The black error
    bars represent statistical uncertainties, and the gray bands total
    point-to-point systematics. A scale uncertainty of 8.3\% from the beam
    polarization measurements is not included. The data are compared to leading
    order Monte Carlo evaluations of $A_{LL}$ assuming various input
    distributions for $\Delta g(x)$ \cite{Jager:2004jh, Gehrmann:1996en,
    deFlorian:2008mr}.}
    \label{fig:all2006}
  \end{figure}

  Figure \ref{fig:all2006} presents STAR's new preliminary results for charged
  pion $A_{LL}$ opposite a trigger jet, plotted versus $z \equiv
  p_{T}(\pi)/p_{T}(jet)$. Full NLO predictions for this observable are in
  preparation \cite{deFlorian:private}; in the interim, the data are compared to
  Monte Carlo evaluations of $A_{LL}$ in which parton distribution functions
  \cite{Jager:2004jh, Gehrmann:1996en, deFlorian:2008mr} that have not been
  ruled out by earlier measurements are sampled at the kinematics specified by
  \textsc{pythia} \cite{Sjostrand:2006za}. Point-to-point systematic
  uncertainties from a variety of sources are summed in quadrature and included
  as the gray bars.
  
  The dominant systematic uncertainty in this analysis arises from the use of
  the JP trigger to select events. This trigger a) hardens the jet $p_{T}$
  spectrum in each z bin, and b) preferentially selects quark jets over gluon
  jets. We quantify a) by using Monte Carlo to determine the trigger efficiency
  for jets at a given $p_{T}$. In the range of jet $p_{T}$ selected for this
  analysis we can parametrize the efficiency as \begin{equation} N_{jets,
  trigger} / N_{jets, total} = 1.149 - 0.2655*p_{T} + 0.01857*p_{T}^{2} -
  0.0003445*p_{T}^{3}.\end{equation} Theoretical predictions for this observable
  should incorporate the trigger efficiency. Future analyses with greater
  statistical precision may be able to avoid this requirement by binning in both
  jet $p_{T}$ and z. For b) we investigate how our LO Monte Carlo evaluation of
  $A_{LL}$ changes when we require the JP trigger condition. Specifically, we
  compare our MC asymmetries for the JP trigger with MC asymmetries that
  incorporate only the trigger \textit{efficiency}. The difference between the
  two is a measure of how the trigger's preference for quark jets affects
  $A_{LL}$.
  
  This contribution presented a ratio of the differential cross sections for
  inclusive charged pion production that shows a clear divergence from unity at
  high $p_{T}$, as well as a new measurement of $A_{LL}$ for charged pion
  production opposite a trigger jet. A forthcoming NLO prediction for this
  observable will allow the measurement to be included in global analyses.
  Comparisons to LO MC evaluations exclude extreme scenarios for the gluon
  distribution function. Future, more precise measurements of $A_{LL}(\pi^{+})$
  have a great potential for providing a better understanding of the gluon
  polarization.

  \bibliographystyle{aipproc}
  
  \bibliography{kocolosk}
  
\end{document}